# Quantum spin Hall states in the lateral heteromonolayers of WTe$_2$–MoTe$_2$

*Mari Ohfuchi and Akihiko Sekine*

Fujitsu Research, Fujitsu Limited, Atsugi, Kanagawa 243-0197, Japan

We used density functional theory to investigate the lateral heteromonolayers of WTe$_2$ and MoTe$_2$. We confirmed that topologically nontrivial and trivial phases are energetically favored for the WTe$_2$ and MoTe$_2$ monolayers, taken out of bulk Td–WTe$_2$ and 2H–MoTe$_2$, respectively. We considered heteromonolayers consisting of these stable building blocks. In the Td–WTe$_2$ and 2H–MoTe$_2$ heteromonolayers with the interfaces oriented perpendicular to the dimer chains of W atoms in Td–WTe$_2$ ($y$ direction), two pairs of helical (quantum spin Hall [QSH]) states, one at each interface, connect the valence and conduction bands. The strain induced by the large lattice mismatch of the two materials in the $y$ direction widens the band gap of the QSH insulator of the Td–WTe$_2$ monolayer and is essential for electronic applications. Furthermore, one-dimensional channels embedded in the layer can help avoid chemical degradation from the edges and facilitate the densification of conducting channels. For the heteromonolayer with interfaces in the $x$ direction, the difference in atomic structure between the two interfaces due to low symmetry creates an energy difference between two helical states and a potential gradient in the wide band gap 2H–MoTe$_2$ region, resulting in various interface localized bands.

INTRODUCTION

Topological insulators (TIs) have been intensively investigated for fundamental research, as well as technological applications, such as spintronics and topological quantum computations.[1–3] Two-dimensional TIs host gapless helical edge states, namely, quantum spin Hall (QSH) states, that preserve time-reversal symmetry.[1,4–7] Monolayer 1T′ transition metal dichalcogenides (TMDCs) are promising candidate materials for QSH insulators.[8,9] Notably, monolayer 1T′–WTe$_2$ is intensively studied because it is the only one where 1T′ is most energetically favored phase under ambient conditions. Localized edge states in the bulk band gap that support the existence of the QSH state have been observed in monolayer 1T′–WTe$_2$.[10–15]

Gapless helical states are usually induced at the interfaces with topological invariant changes,[1,16] such as interfaces between TIs and vacuum and between TIs and normal insulators.[17] In this study, we used density functional theory (DFT) to explore heteromonolayers with interfaces between 1T′–WTe$_2$ as a TI and a topologically trivial TMDC. Helical states have so far been experimentally observed at the interface between the TI domain of 1T′–WSe2 and the normal insulating domain of 1H–WSe2 in mixed-phase WSe$_2$ monolayers.[18] Several heteromonolayers of TMDCs have also been realized experimentally: WS$_2$–MoS$_2$,[19,20] MoSe$_2$–WSe$_2$,[21,22] MoS$_2$–MoSe$_2$,[23] WS$_2$–WSe$_2$,[23] and WSe$_2$–MoS$_2$.[24] Moreover, the one-step controllable growth of the 2H and 1T′ phases of MoTe$_2$ has been demonstrated.[25] The fabrication of the lateral heterostructures of 2H–MoSe$_2$ and 1T′–ReSe$_2$ has also been reported.[26] However, the atomic and electronic structures of 1T′–ReSe$_2$, which is a group VII TMDC, are different from those of group VII TMDCs, such as 1T′–WTe$_2$.[27] Thus, we selected MoTe$_2$ as a candidate material to construct heteromonolayers.

RESULTS AND DISCUSSION

Bulk WTe$_2$ and MoTe$_2$ exhibit Td and 2H crystal structures, respectively.[28–30] We started with the monolayers taken from bulk WTe$_2$ and MoTe$_2$ with Td and 2H crystal structures.[31] We confirmed that the WTe$_2$ and MoTe$_2$ monolayers taken from bulk Td–WTe$_2$ and 2H–MoTe$_2$, respectively, are energetically favored relative to their respective counterparts (Table S1). The optimized lattice constants are within 2% of the experimental values (Table S1).[32–34] We refer to these stable structures as the Td–WTe$_2$ and 2H–MoTe$_2$ monolayers. Notably, the Td–WTe$_2$ monolayer with the optimized atomic structure acquired inversion symmetry and can also be named as the 1T′ monolayer. We considered heteromonolayers consisting of these stable building blocks. The W atoms of the Td–WTe$_2$ monolayer form dimer chains along the $x$ direction. A large lattice mismatch of 3% is observed in the $y$ direction between the two materials, and the lattice mismatch of 0.6% is observed in the $x$ direction (Table S1).

We examined three models of heteromonolayers as shown in Table 1. The structure a20, for example, means that the unit cell is 20 times in the $x$ direction for each Td–WTe$_2$ and 2H–MoTe$_2$ monolayer. We considered two b10 models with different atomic structures around the interfaces. The 20- and 10-unit cells in the $x$ and $y$ directions, respectively, have approximately the same width. Table 1 shows that the formation energies are small and do not significantly depend on the interface direction or the atomic structure of the interface.

**Table 1.** Lattice constants and formation energies of the optimized heteromonolayers of Td–WTe$_2$ and 2H–MoTe$_2$.

| Structure | | a20 | b10-1 | b10-2 |
|---|---|---|---|---|
| Lattice constant (Å) | a | 142.57 | 3.54 | 3.54 |
| | b | 6.26 | 124.96 | 124.91 |
| Formation energy (eV/atom) | | 0.015 | 0.011 | 0.019 |

Figure 1a presents the atomic structure of the a20 heteromonolayer, which has interfaces perpendicular to the W dimer chain in Td–WTe$_2$ ($y$ direction). The band diagram and the spin densities of the wave functions are presented in Figure 1b and 1c, respectively. The bulk band gap appears to lie just below the Fermi level. We can see that two pairs of helical states, that is, QSH states, one at each interface, connect the valence and conduction bands (marked with red broken lines). The localized spin densities spread five or six unit cells from the interface toward the Td–WTe$_2$ monolayer. This finding is consistent with the experimental observation on the interface state decay length of 2 nm into bulk 1T′–WSe$_2$ in mixed-phase WSe$_2$ monolayers.[18] We confirmed that the local density of states (LDOS) of the center in the Td–WTe$_2$ region exhibit the gap below the Fermi level and those at the interface have finite values in the gap (Figure S1).

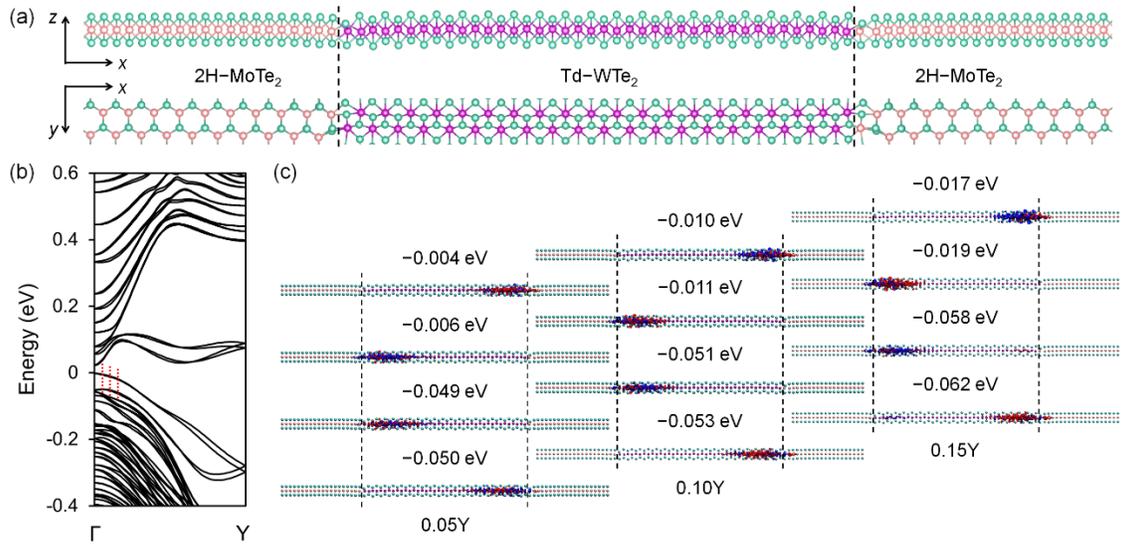

**Figure 1.** Atomic and electronic structures of the lateral heteromonolayer of Td–WTe$_2$ and 2H–MoTe$_2$ with interfaces in the $y$ direction. (a) Side and top views of the atomic structures in the upper and lower panels, respectively. The purple, orange, and green spheres denote W, Mo, and Te atoms, respectively. (b) Energy band diagram drawn with respect to the Fermi level. (c) Spin

densities for the points indicated by the red broken lines in (b). Numerical values show the wavenumber and energy at each point. Blue and red represent up and down spins, respectively.

As mentioned above, a large lattice mismatch exists for the joint direction of the two materials. The lattice constants of the Td–WTe$_2$ and 2H–MoTe$_2$ monolayers are 6.34 and 6.13 Å, respectively. The optimized lattice constant is 6.26 Å. This result indicates that the lattice of Td–WTe$_2$ in the $y$ direction is compressed by 1.3% and that in the $x$ direction is tensile strained. We found that the strain in the $x$ direction is 0.3% in the center of the Td–WTe$_2$ region. Such strain has been reported to widen the band gap of 1T′–WTe$_2$ by several tens of meV in a strain-engineered experiment.[15] A dip near the Fermi level has been observed to become a full gap with zero LDOS with the strain.[15] We also examined the effect of the strain on the band gap by using DFT. Figure 2 presents the band gap as the functions of tensile strain in the $x$ direction and compressed strain in the $y$ direction. Our calculations indicate that the strains of 0.3% and −1.3% in the $x$ and $y$ directions, respectively, widen the band gap by approximately 10 meV. Although this value is smaller than that observed in the experiment, generalized gradient approximation (GGA) is well known to underestimate the band gap in DFT calculations. Our result seems to be consistent with the experimental value. The strain-widened band gap is essential for the electronic applications of the QSH state. Furthermore, one-dimensional channels embedded in the layer can help avoid chemical degradation from the edges and facilitate the densification of conducting channels.

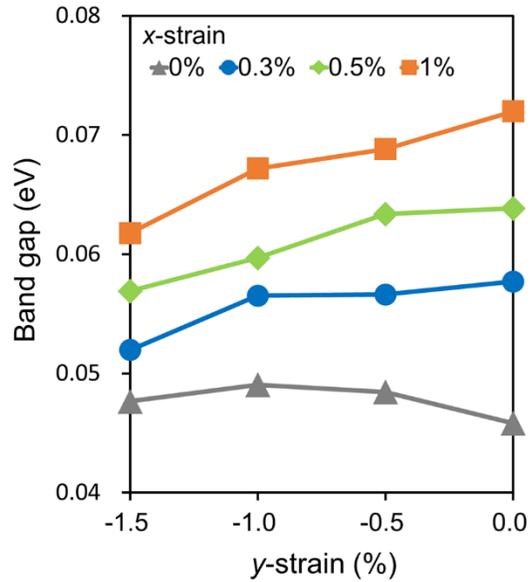

**Figure 2.** Band gap of the Td–WTe$_2$ monolayer as a function of the compressive strain in the *y* direction depending on the tensile strain in the *x* direction.

We move on to lateral heteromonolayers with interfaces in the *x* direction. Figure 3 shows the atomic structures of the b10-1 and b10-2 models. The atomic structures at the right interface are different from each other, as seen in the top views of the atomic structure. One of WTe$_2$ or MoTe$_2$ is selected as the initial atomic structure for structural optimization. As a result, the atomic structure of b10-1 at the right interface is similar to that of WTe$_2$. By contrast, the atomic structure of b10-2 at the right interface is similar to that of MoTe$_2$. We can consider an infinite number of atomic structures, even those that maintain stoichiometry. Although we previously investigated several other atomic structures, we could not find an energetically favorable structure.

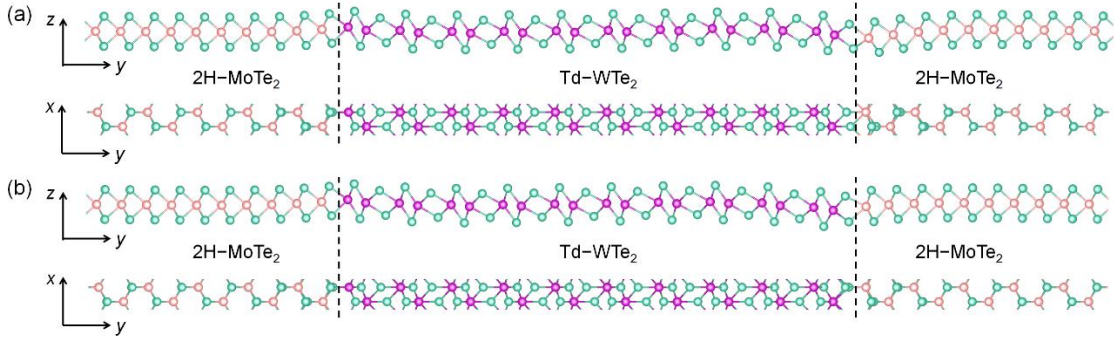

**Figure 3.** Atomic structure of the (a) b10-1 and (b) b10-2 heteromonolayers. The upper and lower panels are the side and top views, respectively, of (a) and (b). The interfaces are along the *x* direction. The purple, orange, and green spheres denote W, Mo, and Te atoms, respectively.

Figure 4a and 4b present the band diagrams of the b10-1 and b10-2 heteromonolayers. We found two pairs of helical states that are similar to the a-20 heteromonolayer. The interface localized bands (marked by red broken lines, see also Figure S2) are highly dependent on the atomic structure of the interface. For the b10-1 and b10-2 heteromonolayers, the spin densities of the upper two bands are localized at the interface on the right, and the lower two bands are localized at the interface on the left. For both models, although the lower two bands lie at approximately 0 eV, the energy positions of the upper two bands are quite different. The upper bands are located at approximately 0.4 and 0.1 eV for b10-1 and b10-2, respectively. For b10-2, other four bands localized at the interface on the right side appear around X points (also marked by a red broken line). Drawing the band structure of b10-1 shows that these bands appear to have fallen off the conduction band. We can confirm these properties in the LDOS in the Td–$WTe_2$ region as shown in Figure 4c and 4d. The energy difference between the two helical states creates the charge distribution in the Td–$WTe_2$ region and a potential gradient in the wide band gap 2H–$MoTe_2$ region (Figure S3). For the b10-1 (b10-2) monolayer, the energy difference is large (small). Likewise, its potential gradient is large (small). These differences appear to have been caused by the low symmetry of the atomic structure, resulting in various interface localized bands.

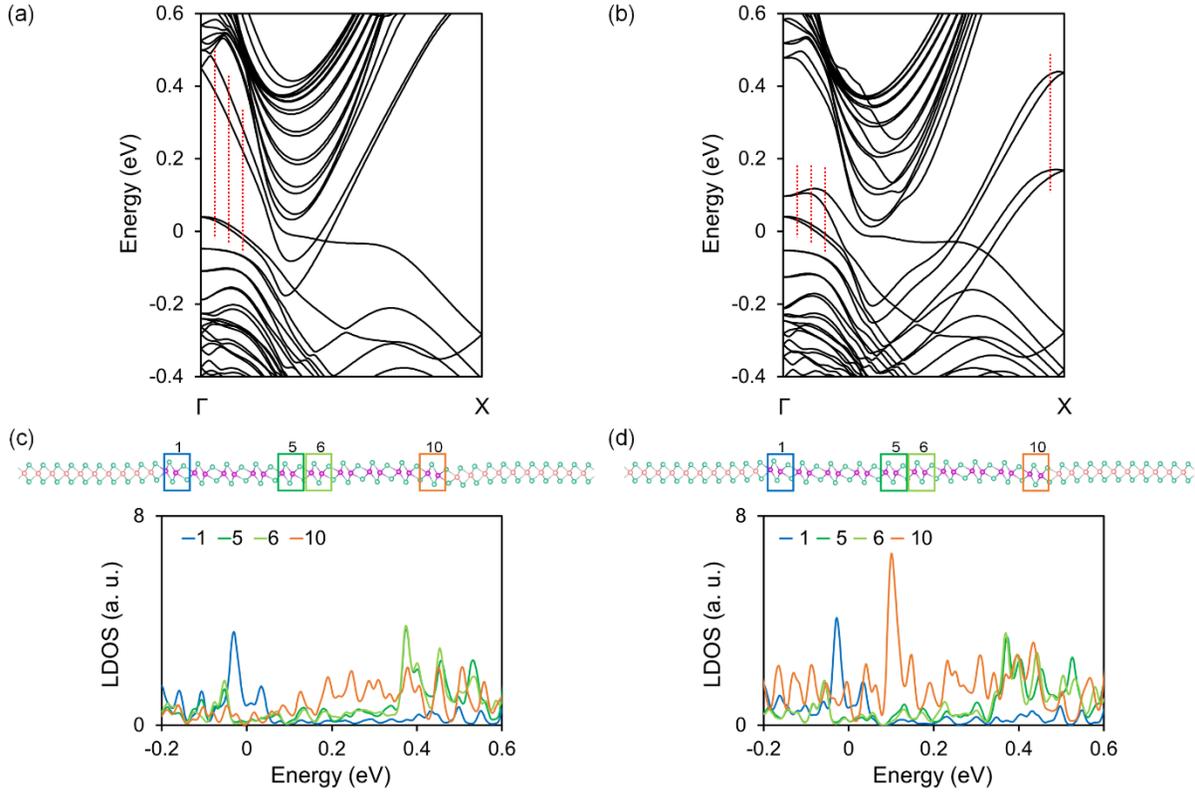

**Figure 4.** (a) (b) Energy band diagrams and (c) (d) LDOS of the b10-1 and b10-2 heteromonolayers, respectively. The energy band diagrams are drawn with respect to the Fermi level. The spin densities are presented for the points indicated by the red broken lines in Figure S2. The areas for the LDOS are indicated in the side view of the atomic structure in the upper panel. The purple, orange, and green spheres denote W, Mo, and Te atoms, respectively.

CONCLUSIONS

We investigated the lateral heteromonolayers of $WTe_2$ and $MoTe_2$ by using DFT. We confirmed that topologically nontrivial and trivial phases are energetically favored for the $WTe_2$ and $MoTe_2$ monolayers. We considered the heteromonolayers consisting of these stable building blocks. For the heteromonolayer of Td–WTe$_2$ and 2H–MoTe$_2$ with interfaces in the $y$ direction, two pairs of helical states, one at each interface, connect the valence and conduction bands. The strain induced by the large lattice mismatch of the two materials in the $y$ direction widens the band gap of the QSH insulator of the Td–WTe$_2$ monolayer. The strain-widened band gap is essential for the electronic applications of QSH states. For the heteromonolayer with interfaces in the $x$ direction, the difference in atomic structure between the two interfaces due to low symmetry creates an energy difference between the two helical states and a potential gradient in the wide band gap 2H–MoTe$_2$ region, resulting in various interface localized bands.

## METHODS

We adopted the OpenMX code[32–34] to perform DFT calculations. This code has been successfully used to study the topological properties of TMDCs and their heterostructures.[18,35–39] The exchange–correlation potential was treated with a GGA.[40] Electron–ion interactions were described by using norm-conserving pseudopotentials.[41,42] The pseudoatomic orbitals denoted by W7.0-s3p2d2f1, Mo7.0-s3p2d2f1, and Te7.0-s3p3d2f1 were used as the basis set. Atomic structure optimizations were performed without considering spin-orbit coupling (SOC).[43,44] The energy bands were calculated with the inclusion of SOC for the final atomic structures.

To investigate the strain effect on the band gap, the lattice constants were determined, then the atomic structure was optimized. The energy bands were calculated with SOC for the atomic structure. The minimum energy gap between the valence and conduction bands was searched in the Brillouin zone.

## ASSOCIATED CONTENT

**Supporting Information**. The following files are available free of charge.

Lattice constant and formation energy of monolayer WTe$_2$ and MoTe$_2$, local density of states (LDOS) for the heteromonolayer with the interfaces in the *y* direction, and spin densities and LDOS for the heteromonolayer with the interfaces in the *x* direction (PDF)

AUTHOR INFORMATION

**Corresponding Author**

Mari Ohfuchi − Fujitsu Research, Fujitsu Limited, Atsugi, Kanagawa 243-0197, Japan; Email: mari.ohfuti@fujitsu.com

**Author**

Akihiko Sekine − Fujitsu Research, Fujitsu Limited, Atsugi, Kanagawa 243-0197, Japan

**Author Contributions**

M.O. carried out the DFT calculations. All authors discussed and commented on the manuscript.

ACKNOWLEDGMENT

We thank the Research Institute for Information Technology, Kyusyu University, Japan for providing access to the supercomputer system for performing the calculations in this study.